\documentclass[10pt,prd,preprintnumbers,floatfix,aps,nofootinbib,showpacs,twocolumn,superscriptaddress,noshowpacs]{revtex4-1} 
\usepackage[latin9]{inputenc}
\usepackage[dvipsnames]{xcolor}
\usepackage{graphicx}
\usepackage{hyperref}
\usepackage{amsthm,amsmath,amssymb,amsfonts}
\usepackage{amsbsy}
\usepackage{bbm}
\usepackage{mathtools}
\usepackage{microtype}
\usepackage{soul}

\usepackage[dvipsnames,svgnames,table]{xcolor}
\usepackage{verbatim}
\usepackage[T1]{fontenc}
\usepackage[latin9]{inputenc}
\usepackage{babel}
\usepackage{float}
\usepackage{graphicx}
\usepackage{svg}
\usepackage{hyperref}
\usepackage{amsthm,amsmath,amssymb,amsfonts}
\usepackage[mathscr]{eucal}
\usepackage{microtype}
\usepackage{esint}
\usepackage{amsbsy}
\usepackage{color}
\usepackage[capitalise]{cleveref}
\usepackage{breakurl}
\usepackage[normalem]{ulem}
\usepackage{bbm}
\usepackage{mathtools}
\usepackage{braket}
\usepackage{comment}
\usepackage[super]{nth}
\usepackage{csquotes}
\usepackage{lmodern}
\usepackage{hyphenat}

\definecolor{lightgray}{gray}{0.9}
\definecolor{Amber}{rgb}{1.0, 0.75, 0.0}
\definecolor{blizzardblue}{rgb}{0.67, 0.9, 0.93}
\definecolor{myblue}{HTML}{0273B2}
\colorlet{myblueLight}{myblue!35}

\makeatletter

\pdfpageheight\paperheight
\pdfpagewidth\paperwidth

\definecolor{lightgray}{gray}{0.9}

\@ifundefined{textcolor}{}{%
 \definecolor{BLACK}{gray}{0}
 \definecolor{WHITE}{gray}{1}
 \definecolor{RED}{rgb}{1,0,0}
 \definecolor{GREEN}{rgb}{0,1,0}
 \definecolor{BLUE}{rgb}{0,0,1}
 \definecolor{CYAN}{cmyk}{1,0,0,0}
 \definecolor{MAGENTA}{cmyk}{0,1,0,0}
 \definecolor{YELLOW}{cmyk}{0,0,1,0}
}

\makeatother

\makeatletter

\DeclareRobustCommand{\rcite}[1]{%
  \rcite@aux#1,\@nil{#1}%
}
\def\rcite@aux#1,#2\@nil#3{%
  \if\relax#2\relax
    Ref.~\cite{#3}%
  \else
    Refs.~\cite{#3}%
  \fi
}
\makeatother
\definecolor{wine}{RGB}{136,34,85}
\definecolor{teal}{RGB}{0,85,102}
\hypersetup{
    colorlinks = true,
    citecolor = {wine},
    linkcolor = {teal},
    urlcolor = {teal},
}

\newcommand{\be}{\begin{equation}}
\newcommand{\ee}{\end{equation}}
\newcommand{\ba}{\begin{eqnarray}}
\newcommand{\ea}{\end{eqnarray}}

\newcommand{\dd}{\mathrm{d}}

\newcommand{\lcdm}{$\Lambda$CDM}
\newcommand{\lameff}{\lambda_{{\rm eff}}}

\newcommand{\nn}{\nonumber}

\def\half{\frac{1}{2}}

\begin{document}

\preprint{IFT-UAM/CSIC-25-77, WUCG-25-12}

\title{DESI constraints on 
two-field quintessence with 
exponential potentials}

\author{George Alestas}
\email{g.alestas@csic.es}
\affiliation{Instituto de F\'isica Te\'orica (IFT) UAM-CSIC, C/ Nicol\'as Cabrera 13-15, Campus de Cantoblanco UAM, 28049 Madrid, Spain}

\author{Marienza Caldarola}
\email{marienza.caldarola@csic.es}
\affiliation{Instituto de F\'isica Te\'orica (IFT) UAM-CSIC, C/ Nicol\'as Cabrera 13-15, Campus de Cantoblanco UAM, 28049 Madrid, Spain}

\author{Indira Ocampo}
\email{indira.ocampo@csic.es}
\affiliation{Instituto de F\'isica Te\'orica (IFT) UAM-CSIC, C/ Nicol\'as Cabrera 13-15, Campus de Cantoblanco UAM, 28049 Madrid, Spain}

\author{Savvas Nesseris}
\email{savvas.nesseris@csic.es}
\affiliation{Instituto de F\'isica Te\'orica (IFT) UAM-CSIC, C/ Nicol\'as Cabrera 13-15, Campus de Cantoblanco UAM, 28049 Madrid, Spain}

\author{Shinji Tsujikawa}
\email{tsujikawa@waseda.jp}
\affiliation{Department of Physics, Waseda University, 3-4-1 Okubo, Shinjuku, Tokyo 
169-8555, Japan}

\date{\today}

\begin{abstract}
We investigate a quintessence model involving two scalar fields with double-exponential potentials. This configuration allows the system as a whole to emulate the dynamics of a single field with a shallower potential, enabling scalar fields that individually cannot drive cosmic acceleration to collectively achieve and sustain it. We assess the viability of this model by performing a fully Bayesian analysis and confronting its predictions with observational data, including the Planck 2018 cosmic microwave background (CMB) shift parameters, the newly released Dark Energy 
Spectroscopic Instrument (DESI) DR2 baryon acoustic oscillation (BAO) measurements, and the Dark Energy Survey Year 5 (DESY5) type Ia supernova (SnIa) sample. 
Our analysis shows that the two-field quintessence model yields a log Bayes factor relative to the flat $\Lambda$ cold dark matter model of $\Delta \ln B \sim 4$, indicating moderate evidence against the latter. We also find that the central values of the two slopes of the exponential potentials are both close to 1, whereas the slope of an effective single-field system is constrained to be less than order unity. This property is theoretically desirable from the perspective of higher-dimensional theories. Thus, the two-field quintessence model with exponential potentials provides a physically motivated and compelling mechanism that is consistent with both observational and theoretical requirements.
\end{abstract}

\maketitle
%

\section{Introduction \label{sec:introduction}}

Understanding the origin of the late-time cosmic acceleration \cite{SupernovaSearchTeam:1998fmf,SupernovaCosmologyProject:1998vns} remains a major unresolved problem in cosmology \cite{Sahni:1999gb,Carroll:2000fy,Peebles:2002gy,Padmanabhan:2002ji,Copeland:2006wr,Weinberg:2013agg,Tsujikawa:2013fta,Joyce:2014kja}. 
Within the framework of general relativity and the standard model of particle physics, the cosmological constant $\Lambda$ can account for this phenomenon. 
However, if $\Lambda$ originates from the vacuum energy associated with standard model particles, 
the theoretically predicted energy scale is vastly larger than the value inferred 
from observations \cite{Weinberg:1988cp,Martin:2012bt}.
Furthermore, the standard $\Lambda$CDM model, which incorporates $\Lambda$ and cold dark matter (CDM) as the dark components of the Universe, exhibits tensions between low- and high-redshift observations \cite{Perivolaropoulos:2021jda}, underscoring the need for extensions to address these discrepancies.

Dark energy (DE), responsible for the late-time acceleration of the Universe, can arise from a dynamical origin, in which its equation of state, 
$w_{\rm DE}$, evolves with time. 
A representative model of dynamical DE is a minimally coupled, canonical scalar field $\phi$, known as quintessence \cite{Fujii:1982ms,Ford:1987de,Ratra:1987rm,Wetterich:1987fm,Chiba:1997ej,Ferreira:1997au,Copeland:1997et,Caldwell:1997ii,Zlatev:1998tr}.
The energy scale of the quintessence potential $V(\phi)$ must be sufficiently low to match the present DE scale, but unlike the cosmological constant, it can increase toward the asymptotic past. Consequently, numerous attempts have been made to construct quintessence models motivated by high-energy physics \cite{Binetruy:1998rz,Brax:1999gp,Copeland:2000vh,Kallosh:2002gf,Choi:1999xn,Nomura:2000yk,Townsend:2001ea,Kim:2002tq,Hall:2005xb,Panda:2010uq}.

In higher-dimensional gravitational theories, such as Kaluza-Klein and supergravity theories, exponential potentials for scalar fields often emerge from the curvature of internal spaces associated with the geometry of extra dimensions \cite{Green:1987sp,Olive:1989nu,Bergshoeff:1996ui}. 
Moreover, exponential potentials can arise in gaugino condensation as a nonperturbative 
effect \cite{deCarlos:1992kox} and in the presence of supergravity corrections to global supersymmetric 
theories \cite{Copeland:2000vh}. 
To realize cosmic acceleration, the slope of 
the potential $V(\phi)=V_0 e^{-\lambda \phi}$ should lie 
within the range $|\lambda|<\sqrt{2}$, where we have adopted units in which the reduced Planck mass 
$M_{\rm pl}$ is set to 
unity. However, the slope of the potential predicted by the aforementioned theories typically falls within the range $|\lambda| \gtrsim 1$, which makes it challenging to realize sufficient late-time cosmic acceleration using a single-field exponential potential. 
In string theory, the swampland conjecture \cite{Obied:2018sgi,Ooguri:2018wrx} 
also states that the slope of the scalar potential must satisfy the bound 
$\left| V_{,\phi} \right / V| >
\mathcal{O}(1)$, 
where $V_{,\phi}=
{\rm d} V/{\rm d} \phi$.

In higher-dimensional theories, multiple scalar fields often appear, 
such as moduli fields associated with the sizes of compactified dimensions.
Even if each individual field has a potential too steep to drive cosmic acceleration on its own, acceleration can still be achieved 
through the collective dynamics of multiple fields.
This mechanism was first recognized in the context of inflation \cite{Liddle:1998jc}, in models involving a sum of exponential potentials of the form $\sum_{i=1} V_i e^{-\lambda_i \phi}$. A more detailed discussion of other two-field models, motivated by heterotic string theory, was presented in Ref.~\cite{Akrami:2018ylq}, while they were also considered in the context of cascade inflation in \cite{Ashoorioon:2006wc,Rezazadeh:2022lsf}, where several steep scalar fields with exponential potentials can be mimicked with a scalar field with a shallower potential. 
In the absence of matter, the multiple fields evolve in such a way that their dynamics mimic those of a single-field model with an effective slope $\lambda_{\rm eff}$ 
satisfying the relation 
$1/\lambda_{\rm eff}^2=
\sum_{i=1} 1/\lambda_i^2$. 
Thus, even if $|\lambda_i| > \sqrt{2}$ for each $i$, the condition for cosmic acceleration, $|\lambda_{\rm eff}| < \sqrt{2}$, can still be satisfied.

In the presence of a barotropic perfect fluid, the mechanism of 
``assisted quintessence'' can also 
operate to drive late-time cosmic 
acceleration through multiple scalar fields with independent exponential potentials \cite{Coley:1999mj,Kim:2005ne,Tsujikawa:2006mw}.
One notable feature of multi-field quintessence is its capacity to induce transient phases of cosmic acceleration \cite{Blais:2004vt,Ohashi:2009xw}. 
This means that the Universe can undergo periods of accelerated expansion that are not permanent. Such dynamics are theoretically important, as they offer 
a possible resolution to the cosmic coincidence problem---that is, why the energy densities of DE and matter are of the same order of magnitude today, despite their different evolutionary histories over cosmic time \cite{Zlatev:1998tr}. 
In models where cosmic acceleration is a transient phase, preceded by scaling radiation and matter 
eras \cite{Ohashi:2009xw}, the present epoch does not require fine-tuning of parameters to sustain a constant DE density indefinitely, thereby alleviating the coincidence problem. 

Recent baryon acoustic oscillation (BAO) measurements from the Dark Energy Spectroscopic Instrument (DESI) \cite{DESI:2024mwx,DESI:2025zgx,Lodha:2025qbg} 
suggest that dynamical DE models are statistically 
favored over the cosmological constant.  
By adopting a parametrization of the DE equation of state $w_{\rm DE}$, joint analyses of DESI data combined with various SnIa datasets indicate a preference for dynamical DE at a significance level ranging from $2.8\sigma$ to $4.2\sigma$, depending on the specific SnIa dataset used. 

For the single-field exponential potential 
$V(\phi)=V_0\,e^{-\lambda \phi}$, 
the DESI data constrain the slope 
parameter to  
$\lambda = 0.698^{+0.173}_{-0.202}$ at the $1\sigma$ level, assuming a spatially-flat Friedmann-Lema\^{i}tre-Robertson-Walker (FLRW) background \cite{Akrami:2025zlb}. 
This constraint remains robust even when spatial curvature is taken into account~\cite{Alestas:2024gxe, Bhattacharya:2024hep}. Moreover, related analyses of other types of quintessence potentials, such as Higgs-like potentials, axion-like thawing models, and thawing quintessence in general \cite{Gialamas:2025pwv, Tada:2024znt, Wolf:2024eph, Lin:2025gne}, show that quintessence remains viable at a similar level of significance when the latest cosmological data are used.

The main advantage of employing physically motivated theories to model the background expansion of the Universe, rather than using Taylor expansions of DE at late times or other ad hoc parameterizations, is that one avoids the artificial introduction of information not present in the data. Such artificially introduced information has been shown to lead to markedly different and potentially misleading conclusions \cite{Nesseris:2025lke}.
Indeed, when theory-informed priors are applied in fitting these Taylor expansions, the evidence for dynamical DE decreases from approximately 
$3.1\sigma$ to $1.3\sigma$ \cite{Toomey:2025xyo}. Finally, several analyses have explored the possibility of systematic effects in the DESI data, which also affect the evidence supporting 
evolving DE \cite{Sapone:2024ltl, Colgain:2024mtg, Chudaykin:2024gol, Colgain:2024xqj}, or various extended models and data combinations  \cite{RoyChoudhury:2024wri,RoyChoudhury:2025iis, RoyChoudhury:2025dhe}.

In this paper, we aim to constrain the two-field quintessence model with exponential potentials by confronting it with DESI data in combination with other observational datasets, and to assess its viability as an alternative to the $\Lambda$ cold 
dark matter ($\Lambda$) model.

To place observational constraints on the model parameters, we integrate the background equations of motion without 
resorting to specific DE parameterizations. 
We investigate how each slope, $\lambda_i$, of the exponential potentials is
constrained by the DESI data.
In particular, we are interested in whether $\lambda_i$ can be of order unity, as required from the theoretical perspective. 
Thus, by adding one more scalar field, our aim is not to improve the fit to the data, but rather to address the theoretical question of whether both the swampland conjectures and the observations can be satisfied simultaneously, which is not possible with a single field. We discuss our findings in the following sections.

Specifically, our paper is organized as follows. 
In Sec.~\ref{sec:theory}, we introduce 
the two-field quintessence model and revisit its dynamics. 
In Sec.~\ref{sec:data}, we describe the observational datasets used to constrain the models, as well as the statistical methods employed for their comparison.
In Sec.~\ref{sec:results}, we present the results of a 
a Markov chain Monte Carlo (MCMC) analysis, including constraints on the model parameters and a comparison of their goodness of fit to the data using Bayesian model selection. 
In Sec.~\ref{sec:conclusions}, we conclude with a concise summary of our main findings.

\section{Theoretical formalism}
\label{sec:theory}

We consider the cosmological dynamics of two canonical scalar fields, $\phi$ and $\chi$, which are minimally coupled to gravity and serve as components of DE, in the presence of nonrelativistic matter (including baryons 
and CDM) and radiation~\cite{Liddle:1998jc,Malik:1998gy,Tsujikawa:2006mw,Ohashi:2009xw,vandeBruck:2009gp}. The system is described by the following action
\ba
{\cal S} &=& \int 
\dd^4 x\,\sqrt{-g}\,\Bigg[ 
\frac{1}{2}R - \frac{1}{2} g^{\mu \nu} \partial_{\mu}\phi \partial_{\nu}\phi - \frac{1}{2} g^{\mu \nu} \partial_{\mu}\chi \partial_{\nu}\chi \nn \\
&&-V(\phi, \chi) + \mathcal{L}_{\rm{m}}
+\mathcal{L}_{\rm{r}} \Bigg]\,,
\label{action}
\ea
where $g$ is the 
determinant of the metric tensor 
$g_{\mu \nu}$, $R$ is the Ricci scalar, 
$V$ is the potential of two scalar 
fields $\phi$ and $\chi$, 
and $\mathcal{L}_{\rm{m}}$ and 
$\mathcal{L}_{\rm{r}}$
denote the Lagrangian densities of nonrelativistic matter and radiation, respectively. The background geometry is assumed to follow the FLRW metric:
\begin{equation}
\dd s^2=-\dd t^2 + a^2(t) \left( \frac{\dd r^2}{1-kr^2} + r^2 \dd\theta^2 +r^2\sin{\theta}^2 \dd\phi^2 \right)\,,
\end{equation}
where $a(t)$ is the scale factor and $k \in {-1,0,+1}$ specifies the spatial curvature, corresponding respectively to a homogeneous and isotropic Universe with negative (hyperbolic), zero (flat), or positive (spherical) curvature.
In the matter sector, we consider pressureless matter with energy density $\rho_{\rm m}$, and radiation with energy density $\rho_{\rm r}$ and pressure $p_{\rm r} = \rho_{\rm r}/3$.

The evolution of the background is governed by the Friedmann equations:
\begin{align}
3 H^2 &=
\rho_{\rm DE}+\rho_{\rm m}+\rho_{\rm r}-\frac{3k}{a^2}\,,
\label{eq:Friedmann:scalar:one}\\
2 \dot{H} &= -\rho_{\rm DE}-p_{\rm DE}-\rho_{\rm m}-\rho_{\rm r}-p_{\rm r}+\frac{2k}{a^2}\,,\label{eq:Friedmann:scalar:two}
\end{align}
where $H \equiv \dot{a}/a$ is the Hubble parameter, an overdot denotes a derivative with respect to the cosmic time $t$, and 
\ba
\rho_\mathrm{DE}&=&\half \dot{\phi}^2 +\half\dot{\chi}^2+V(\phi,\chi)\,,\label{eq:de_density}\\
p_\mathrm{DE}&=&\half\dot{\phi}^2 +\half\dot{\chi}^2 -V(\phi,\chi)\,,
\ea
which correspond to the DE density and pressure, respectively. 
One can express Eq.~(\ref{eq:Friedmann:scalar:one}) 
in the form 
\be
1=\Omega_{\rm DE}
+\Omega_{\rm m}+\Omega_{\rm r}
+\Omega_{k}\,,
\ee
where $\Omega_{\rm DE}=\rho_{\rm DE}/(3H^2)$, 
$\Omega_{\rm m}=\rho_{\rm m}/(3H^2)$, 
$\Omega_{\rm r}=\rho_{\rm r}/(3H^2)$, and 
$\Omega_k=-k/(a^2 H^2)$ are 
the density parameters of DE, nonrelativistic matter, radiation, and spatial curvature, respectively. We define the DE equation of state, as
\begin{equation}
w_\mathrm{DE}=
\frac{p_\mathrm{DE}}{\rho_\mathrm{DE}}=
\frac{\dot{\phi}^2/2 
+\dot{\chi}^2/2 -V(\phi,\chi)}{\dot{\phi}^2/2 
+\dot{\chi}^2/2 +V(\phi,\chi)}\,.\label{eq:wDE}
\end{equation}
We will focus on a positive potential 
$V(\phi, \chi)$, in which case 
$w_{\rm DE}$ is in the range $w_{\rm DE} 
\geq -1$.
Each scalar field obeys the 
following equation of motion:
\ba
\ddot{\phi} + 3H\dot{\phi} 
+V_{,\phi}  =0\,,
\label{eq:continuity:scalar1} \\
\ddot{\chi} + 3H\dot{\chi} 
+ V_{,\chi} =0\,,
\label{eq:continuity:scalar2}
\ea
where $V_{,\phi}=
\partial V/\partial \phi$ and 
$V_{,\chi}=
\partial V/\partial \chi$. Using these equations, the DE 
sector obeys the continuity equation 
$\dot{\rho}_{\rm DE}+3H
(\rho_{\rm DE}+p_{\rm DE})=0$, 
while the matter sector satisfies
the continuity equations 
$\dot{\rho}_{\rm m}+3H \rho_{\rm m}=0$ and $\dot{\rho}_{\rm r}
+4H \rho_{\rm r}=0$. 
We also introduce the total effective equation of state of the Universe, as \cite{Amendola:2015ksp}
\be
w_\mathrm{eff} = -1-\frac23 \frac{\dot{H}}{H^2}\,,
\label{eq:weff} 
\ee
which provides information on the transition epochs of the Universe. In particular, 
$w_{\rm eff} < -1/3$ corresponds to accelerated expansion ($\ddot{a} > 0$).

For concreteness, we consider a potential given by the sum of two exponential terms of the form 
\be
V(\phi,\chi)=
V_{0\phi}\,e^{-\lambda_{\phi}\phi}
+V_{0\chi}\,e^{-\lambda_{\chi}\chi}\,,
\ee
where $V_{0\phi}$ and 
$V_{0\chi}$ are positive constants. 
An attractive feature of this potential is that, in the absence of matter, the two-field system can effectively behave as a single field with an effective potential that is shallower than those of the individual fields. 
Indeed, the above two-field system possesses a scalar-field-dominated fixed point ($\Omega_{\rm DE}=1$) 
with \cite{Coley:1999mj,Kim:2005ne,Tsujikawa:2006mw}
\be
w_{\rm DE}=-1+\frac{\lambda_{\rm eff}^2}{3}\,,
\ee
where the effective slope $\lambda_{\rm eff}$ satisfies
\be 
\frac{1}{\lameff^2} = \frac{1}{\lambda_{\phi}^2}+\frac{1}{\lambda_{\chi}^2}\,.
\label{eq:lameff}
\ee
As long as $\lambda_{\rm eff}^2 < 2$, the condition $w_{\rm DE} < -1/3$ is satisfied, and hence the Universe can enter a phase of cosmic acceleration, approaching the fixed point characterized by DE dominance.
This assisted quintessence mechanism allows scalar fields that are individually unable to drive cosmic acceleration to cooperate and collectively sustain accelerated expansion. 

If one of the scalar fields, say $\phi$, has a slope $\lambda_{\phi}$ larger than order unity, while that of the other field satisfies $\lambda_{\chi} < \mathcal{O}(1)$, there exists a scaling solution characterized by the density parameter $\Omega_{\phi} = 4/\lambda_{\phi}^2$ during the radiation era \cite{Copeland:1997et,Barreiro:1999zs}. This solution is stable for $\lambda_{\phi}^2 > 4$, 
but unstable for $\lambda_{\phi}^2 < 4$. 
Under the existence of a stable scaling solution during the radiation era, the Big Bang nucleosynthesis (BBN) constraint imposes $\Omega_{\phi} < 0.045$ \cite{Bean:2001wt}, which translates to $\lambda_{\phi} > 9.4$. This epoch can be followed by a scaling matter era characterized by $\Omega_{\phi} = 3/\lambda_{\phi}^2$ and the field equation of state $w_{\phi} = 0$.
If the slope of the other field satisfies $\lambda_{\chi} \lesssim {\cal O}(1)$, the system can enter an epoch of cosmic acceleration. 
Indeed, it is also possible to realize transient acceleration for $\lambda_{\phi} \gtrsim 10$ and $\lambda_{\chi} > \sqrt{2}$ \cite{Ohashi:2009xw}. More conditions for which scaling cosmologies are related to late-time attractors of multi-field exponential potentials were derived in Ref.~\cite{Shiu:2023fhb}.

However, the presence of the scaling radiation and matter eras implies that $w_{\rm DE}$ evolves from $1/3$ to $0$ and eventually reaches the regime $w_{\rm DE} < -1/3$ at low redshifts (see, e.g., Fig.~4 in Ref.~\cite{Ohashi:2009xw}). This behavior is generally disfavored by recent observations due to the large deviation of $w_{\rm DE}$ from $-1$ in the region $w_{\rm DE} > -1$ during 
most of the cosmological epoch.
In our statistical analysis, we do not exclude the possibility that one of the potential slopes exceeds order unity. However, unless both $\lambda_{\phi}$ and $\lambda_{\chi}$ are at most of order 1, we will see that the two-field quintessence model is not compatible with the observational data.

\begin{figure}[!t]
\includegraphics[width=0.99\columnwidth]{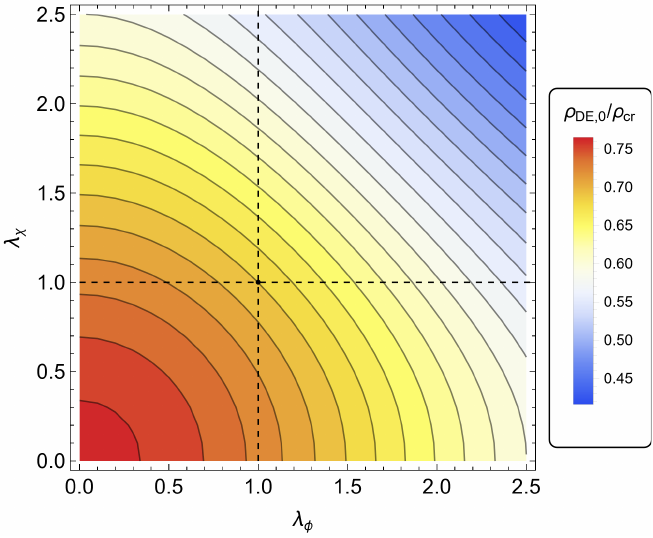}
\caption{The theoretically derived present-day fractional DE density, 
$\rho_{{\rm DE},0}/\rho_\mathrm{cr}$, 
is shown as a function of $\lambda_{\phi}$ 
and $\lambda_{\chi}$ for the 
mean parameter values reported in Table~\ref{tab:MCMC_bf}, 
without performing shooting. 
The initial conditions are chosen such that $\phi$, $\chi$, and their first derivatives are set to 0.
We observe that values of $\lambda_\phi \sim \lambda_\chi \sim 1$ can jointly yield 
$\rho_{{\rm DE},0}/\rho_\mathrm{cr} \sim 0.685$, in agreement with current observational data. Note that $\rho_{{\rm DE},0}/\rho_\mathrm{cr}$ is not necessarily equal to $\Omega_{{\rm DE},0}$ unless a shooting procedure is performed.
\label{fig:theory} }
\end{figure}

In Fig.~\ref{fig:theory}, we show the theoretical contours 
of today's DE density
$\rho_{{\rm DE},0}$, relative to the critical density, 
$\rho_{\rm cr}$, 
as a function of $\lambda_\phi$ and $\lambda_\chi$. Different colors indicate different values of $\rho_{{\rm DE},0}/\rho_{\rm cr}$.
There is a tendency for $\rho_{{\rm DE},0}/\rho_{\rm cr}$ to decrease as $\lambda_\phi$ and $\lambda_\chi$ increase. 
Notably, when both couplings 
are of order unity ($\lambda_\phi \sim \lambda_\chi \sim 1$), the model predicts a present-day DE density parameter of $\Omega_{\rm DE,0} \sim 0.68$, in agreement with current observational constraints. 
Such consistency is generally absent in single quintessence scenarios~\cite{Alestas:2024gxe,Akrami:2025zlb}.

It should also be stressed that, while the qualitative analysis shown in Fig.~\ref{fig:theory} is instructive, the final observationally constrained parameter contours, presented in Sec.~\ref{sec:results}, do not exactly follow this shape due to data uncertainties and the flexibility of the parameterization.

\section{Observational 
data and methodology
\label{sec:data}}

We test our two-field quintessence model against recent observational data.
Specifically, we use the Planck 2018 CMB shift parameters \cite{Planck:2018vyg} and the newly released DESI DR2 BAO measurements~\cite{DESI:2025zgx,Lodha:2025qbg} (for an alternative study on BAO measurements, see Refs.~\cite{Anselmi:2022exn, ODwyer:2019rvi, Anselmi:2018vjz}). 
The CMB distance priors have been shown to provide constraints 
on cosmological parameters that are in good agreement with 
those obtained from the full Planck 2018 data release \cite{Chen:2018dbv,Nesseris:2019fwr,Zhai:2018vmm}.
Lastly, we also include the DESY5 SnIa data in the analysis \cite{DES:2024jxu}.

The CMB distance priors are characterized by three key quantities. The first is the acoustic scale, which determines the angular size of temperature fluctuations in the transverse direction. The second is the shift parameter, which encodes the relative positions of the temperature peaks in the CMB power spectrum. The third is the physical baryon density, defined as $\omega_\mathrm{b} = \Omega_\mathrm{b,0} h^2$. 
Here, $\Omega_{\mathrm{b},0}$ denotes the present-day baryon density parameter, and 
$h$ is defined as
$h = H_0 / (100\,\mathrm{km}\,
\mathrm{s}^{-1}\,
\mathrm{Mpc}^{-1})$,
where $H_0$ is the Hubble expansion rate today. 
For further details, the reader is referred to Ref.~\cite{Zhai:2018vmm}.

\begin{table}
\renewcommand{\arraystretch}{1.3}
\rowcolors{1}{}{lightgray}
    \centering
    \setlength\tabcolsep{0pt}
    \begin{tabular}{ |c|c| }
    \hline
    \rowcolor{myblueLight}
    ~Parameters~ & ~Prior range~ \\
    \hline
    $\Omega_\mathrm{m,0}$  & ~~$\mathcal{U}\left[0, 1\right]$~~ \\
    \hline
    $h$  & ~~$\mathcal{U}\left[0.5, 1\right]$~~ \\
    \hline
    $\lambda_\phi$  & ~~$\mathcal{U}\left[0, 3\right]$~~ \\
    \hline
    $\lambda_\chi$  & ~~$\mathcal{U}\left[0, 3\right]$~~ \\
    \hline
    \end{tabular}
    \caption{The prior ranges used in our analysis, for the parameter vectors as discussed in the text. In all cases we assume a flat uniform prior in the range $\in [a,b]$, denoted as $\mathcal{U}\left[a, b\right]$ in the Table.}
    \label{tab:priors}
\end{table}

To incorporate the BAO data in our analysis, we adopt the approximate expression for the sound horizon at the drag epoch proposed in Ref.~\cite{Brieden:2022heh}, following the methodology of Ref.~\cite{DESI:2024mwx}.
In addition, we impose a BBN  prior on the baryon density, 
$\omega_b = 0.02218 \pm 0.00055$ \cite{Schoneberg:2024ifp}. Under these assumptions, the analysis includes a total of $N = 1845$ data points.

\begin{table*}
\renewcommand{\arraystretch}{1.3}
\rowcolors{1}{}{lightgray}
    \centering
    \setlength\tabcolsep{0pt}
    \begin{tabular}{ |c|c|c|c|c| }
    \hline
    \rowcolor{myblueLight}
    ~Parameters/Quantities~ &  ~~$\Lambda$CDM~~ & 
    ~~$\phi\chi$CDM~~ & 
    ~~CPL~~  \\
    \hline
    $\Omega_{\rm{m},0}$ & ~~$0.305\pm 0.003$~~ & ~~$ 0.315\pm 0.005$~~ & $0.320\pm 0.006$ \\
    $h$& $0.680\pm 0.002$ & $ 0.667\pm 0.005$ & $0.667\pm 0.057$\\
    $\lambda_{\phi}$ & $\cdots$ & $ 1.026\pm 0.547 $ & $\cdots$\\
    $\lambda_{\chi}$ & $\cdots$ & $ 0.962\pm 0.546 $ & $\cdots$\\
    $\lambda_{\rm{eff}}$ & $\cdots$ & $ 0.488\pm 0.255 $ & $\cdots$\\
    $w_0$ & $-1$ & $\cdots$& ~ $-0.751\pm 0.058$\,\\
    $w_a$ & 0 & $\cdots$& ~$-0.877\pm 0.231$\\
    $\chi^2$ & $1681.04$ & $1674.27$ & 1660.65\\
    $\Delta\ln B$ & 0 & 3.97 & 6.84 \\
    \hline
    \end{tabular}
    \caption{This table presents the results of our Bayesian analysis, the marginalized posterior means and $68.3\,\%$ credible intervals on the model parameters, obtained from MCMC scans of the parameter space for the models and parametrizations considered in this work. In the Table, $\phi\chi$CDM represents the two-field quintessence model given by Eq.~\eqref{action}, while CPL corresponds to the parameterization of the DE equation of state in Eq.~\eqref{eq:CPL}. Note that the value of $\lambda_{\rm eff}$ is a derived parameter obtained by directly sampling from the posteriors of $\lambda_\phi$ and $\lambda_\chi$, and therefore differs from the value $\lambda \sim 0.7$ of a single scalar field (see Ref.~\cite{Akrami:2025zlb}), since Eq.~\eqref{eq:lameff} is non-linear. 
    Using instead the mean values of $\lambda_\phi$ and $\lambda_\chi$ directly in Eq.~\eqref{eq:lameff}, we indeed find $\lambda_{\rm eff} \sim 0.7$.
    We also show the corresponding minimum $\chi^2$ values and $\Delta \ln B$, where 
    $\Delta\ln B_X = \ln B_X - \ln B_{\Lambda \mathrm{CDM}}$ for 
    $X=\phi \chi$CDM and 
    $X={\rm CPL}$.
    }
    \label{tab:MCMC_bf}
\end{table*}

\begin{figure*}[!t]
\includegraphics[width=0.95\columnwidth]{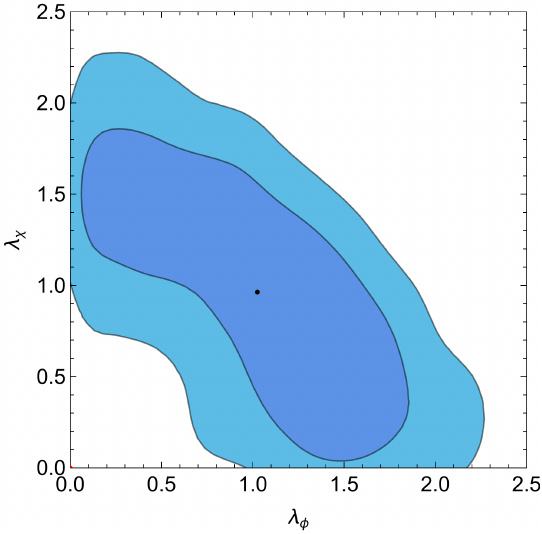}
\includegraphics[width=0.955\columnwidth]{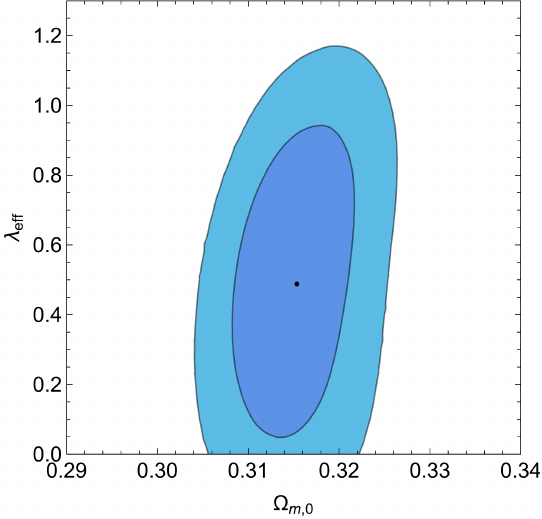}
\caption{
Confidence regions for the parameters $\lambda_{\phi}$ and $\lambda_{\chi}$ (left panel), and for $\lambda_\mathrm{eff}$ versus today's matter density parameter, $\Omega_{\mathrm{m},0}$ (right panel). The constraints are derived from CMB, DESI DR2, and DESY5 data.  
The inner and outer contours correspond to the 68.3\,\% and 95.5\,\% credible intervals, respectively.
\label{fig:cons_lx_lf_and_leff_Om}
}
\end{figure*}

\begin{figure}[!t]
\includegraphics[width=0.99\columnwidth]{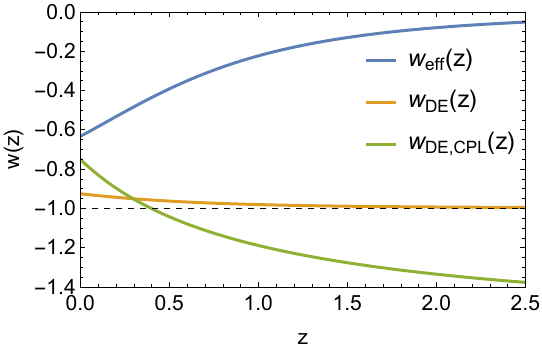}
\caption{Plots of the DE equation of state $w_\mathrm{DE}(z)$ and the total effective equation of state $w_\mathrm{eff}(z)$ are shown for the two-field quintessence model of Eq.~(\ref{action}). 
We also show the evolution of $w_{\rm DE,CPL}(z)$ in the Taylor-expanded CPL model of Eq.~\eqref{eq:CPL}. All plots use the mean parameter values listed in Table~\ref{tab:MCMC_bf}. For the two-field quintessence model, we have $w_\mathrm{eff} = -1/3$ at 
$z=0.65$.
\label{fig:wz_plots}
}
\end{figure}

We also assume the amplitude of each term of the potential is the same, i.e., $V_{0\phi}=V_{0\chi}=V_0$. 
As long as the ratio $r\equiv V_{0\phi}/V_{0\chi}$ does not significantly differ from 1, i.e., both scalar fields contribute equally in the DE energy budget, we confirmed that the results are 
insensitive to $r$. 
In Appendix~\ref{sec:app2}, we explicitly demonstrate that this 
choice is robust, as even sizable deviations of $r$ from 
unity---of order $\sim 20\%$ or larger---do not significantly 
affect the expansion history.

To satisfy the consistency relation 
$H(a=1) = H_0$, we numerically determine the parameter $V_0$ using a shooting method. 
Without loss of generality, we set the initial values of $\phi$, $\chi$, and their first time derivatives to 0 in the radiation era, at approximately $N_{\mathrm{ini}}=-15$, where $N=\ln a$ denotes the e-folding number.
Thus, the final parameter vector for \lcdm{} is $\theta_{\Lambda,i} = 
\{\Omega_{\mathrm{m},0}, h\}$, whereas 
for the two-field quintessence model it is $\theta_{\mathrm{tq},i} = 
\{\Omega_{\mathrm{m},0}, h, \lambda_\phi, \lambda_\chi\}$, and these parameters are varied in our analysis, with the actual priors shown in Table~\ref{tab:priors}.

For comparison, we also perform an analysis using the parametrization of $w_{\rm DE}$ based on its Taylor expansion around $a=1$, given by \cite{Chevallier:2000qy, Linder:2002et}
\be 
w_\mathrm{DE,CPL}(a)=
w_0+w_a\,(1-a)\,,
\label{eq:CPL}
\ee
where $w_0$ and $w_a$ are parameters characterizing the present value of $w_{\rm DE}$ and its derivative, respectively.
Since we do not restrict the parameter space of $w_0$ and $w_a$, the region with $w_{\rm DE}(a) < -1$ is allowed in the CPL parametrization (\ref{eq:CPL}). In contrast, in our two-field quintessence model, $w_{\rm DE}(a)$ is always constrained to satisfy $w_{\rm DE}(a) \geq -1$. 

Finally, it should be stressed that caution is warranted when such Taylor expansions for the DE equation-of-state parameter, as in Eq.~\eqref{eq:CPL}, are used as they may bias the analysis and the resulting conclusions, even if they seem to present a better fit to the data \cite{Nesseris:2025lke}.

We conduct a full Bayesian analysis of both the two-field quintessence model and the $\Lambda$CDM model to constrain their parameters and assess their goodness of fit, assuming a spatially flat Universe ($\Omega_\mathrm{k,0}=0$) 
in both cases.  
To further evaluate the fit of these cosmological models to the data, 
we employ the \texttt{MCEvidence} tool \cite{Heavens:2017afc}, 
which computes the Bayes factors using a nearest neighbor approach.
To validate the robustness of \texttt{MCEvidence}, we also performed a brute-force calculation of the evidence, by numerically evaluating the multidimensional integral, finding excellent agreement between the two approaches. 

Finally, as mentioned in Ref.~\cite{Heavens:2017afc}, the log evidence, i.e., the logarithm of the marginal likelihood, has a theoretical uncertainty given by
$$
\sigma_{\ln B}=\frac{\sigma_E}{E}=\frac{1}{\sqrt{N k+1}},
$$
where $E$ is the evidence, $k$ denotes the $k$-th nearest neighbor of an MCMC sample, and $N$ is the number of MCMC samples. In our case, we take $k=1$, and our MCMC chains contain $N\sim {\cal O}(10^5)$ points. We therefore find $\sigma_{\ln B}\sim 3\times 10^{-3}$. We note that this uncertainty corresponds only to the theoretical uncertainty of the nearest-neighbor algorithm used in the \texttt{MCEvidence} tool \cite{Heavens:2017afc}. 

The interpretation of these Bayes factors follows both the original Jeffreys' scale \cite{Jeffreys:1939xee, Trotta:2008qt} and its more recent refinement \cite{John:2002gg, Nesseris:2012cq}, which associate specific threshold values with qualitative levels of model preference \cite{Trotta:2008qt}.
According to the revised scale, a difference in log Bayes factors, defined as 
$\Delta\ln B_{X,Y} \equiv \ln B_X - \ln B_Y$, between two 
models $X$ and $Y$ implies the following: if $\Delta \ln B_{X,Y} < 1.1$, the evidence does not significantly favor either model. Values within the range $1.1 < \Delta \ln B_{X,Y} < 3$ indicate weak support for model $X$ relative to model $Y$, while 
$3 < \Delta \ln B_{X,Y} < 5$ corresponds to moderate evidence. A value of $\Delta \ln B_{X,Y} > 5$ denotes a strong preference for model $X$.

\section{Bayesian Analysis Results \label{sec:results}}

The main results of our MCMC analysis are presented in Table~\ref{tab:MCMC_bf} and Fig.~\ref{fig:cons_lx_lf_and_leff_Om}. 
Specifically, Table~\ref{tab:MCMC_bf} 
presents the marginalized posterior means and $68.3\%$ credible intervals of the free parameters of three models: $\Lambda$CDM, 
the two-field quintessence (``$\phi \chi$CDM''), and the CPL 
parametrization of $w_{\rm DE}(z)$. 

Since we are interested in performing a fully Bayesian analysis, we focus on the posterior mean values of the parameters rather than their best-fit values. 
In Bayesian inference, the posterior is the relevant quantity, as it represents our state of knowledge about the parameters after incorporating the information from the data \cite{Trotta:2008qt}. Finally, we also show the best-fit $\chi^2$ values, as well as the differences in the log Bayes factors relative to the $\Lambda$CDM model.

We find the mean values of the potential slope parameters to be $\lambda_{\phi} = 1.026$ and $\lambda_{\chi} = 0.962$. This result is consistent with expectations from higher-dimensional theories, which generally predict slopes of order 1. At the 68.3\,\% credible interval, the marginalized posterior mean values of $\lambda_{\phi}$ and $\lambda_{\chi}$ are found to be
\be
\lambda_\phi \sim \lambda_\chi \sim 1 \pm 0.55\,.
\ee

As we observe in the left panel of Fig.~\ref{fig:cons_lx_lf_and_leff_Om}, the model with $\lambda_{\phi}=0$ and $\lambda_{\chi}=0$, which  corresponds to the $\Lambda$CDM limit, lies outside the 95.5\,\% confidence level region. 
The case in which either $\lambda_{\phi}$ or $\lambda_{\chi}$ exceeds 2 is disfavored by the data. This stems from the emergence of scaling radiation and matter eras prior to cosmic acceleration. In such models, the DE equation of state exhibits a large deviation from
$w_{\rm DE}=-1$ during most of the cosmological epoch. 
Therefore, scenarios with $\lambda_{\phi}>2$ or $\lambda_{\chi}>2$ are incompatible with observational constraints.

At this point, we note that, as can be seen from the left panel of Fig.~\ref{fig:cons_lx_lf_and_leff_Om}, the MCMC chains do not exhibit signs of poor stability or lack of convergence. In principle, adopting a narrower prior range for the parameters $\lambda_{\phi}$ and $\lambda_{\chi}$ could reduce the extent of the correlation observed in the posterior; however, this cannot be determined \emph{a priori}. The observed degeneracy instead reflects an exact symmetry of the model rather than a convergence issue. Imposing a prior that enforces an ordering between the two parameters would merely remove a duplicate region of parameter space without introducing additional physical information. Although such a choice could improve sampling efficiency and speed up convergence by roughly a factor of two, it is not required for the present analysis and might be useful for future analyses.

In the right panel of Fig.~\ref{fig:cons_lx_lf_and_leff_Om}, we find that the mean value of the effective slope is constrained to be $\lambda_\mathrm{eff} \sim 0.5 \pm 0.25$ at the 68.3\,\% 
credible interval. 
The reason why this value differs from that obtained by simply substituting the mean values of $\lambda_\phi$ and $\lambda_\chi$ into Eq.~\eqref{eq:lameff} is that the mean value of $\lambda_{\rm eff}$ computed by sampling the posterior distributions of the slope parameters does not, in general, coincide with the value obtained by directly using their means.
This can be written explicitly as
\be
\langle \lambda_{\rm eff}(\lambda_\phi,\lambda_\chi) \rangle \ne  \lambda_{\rm eff}(\langle \lambda_\phi \rangle,\langle \lambda_\chi \rangle)\,,
\ee
since the transformation in Eq.~\eqref{eq:lameff} is non-linear and the 1D posterior distributions of $\lambda_\phi$ and $\lambda_\chi$, shown in Appendix~\ref{sec:app1}, are noticeably 
non-Gaussian.

Thus, since the value of $\lambda_{\rm eff}$ is obtained by directly sampling the posterior distributions of $\lambda_\phi$ and $\lambda_\chi$, its mean reported in Table~\ref{tab:MCMC_bf} differs from that 
of a single scalar field, $\lambda \sim 0.7$ \cite{Akrami:2025zlb}. If, instead, one substitutes the mean values of $\lambda_\phi$ and $\lambda_\chi$ directly into Eq.~\eqref{eq:lameff}, one obtains $\lambda_{\rm eff} \sim 0.7$. 
In any case, owing to the cooperative dynamics of the two scalar fields, the mean value of the effective slope, $\lambda_{\rm eff}$, is reduced to below unity.

In Fig.~\ref{fig:wz_plots}, we plot the evolution of $w_{\rm DE}$ and $w_{\rm eff}$
as functions of redshift, 
$z = 1/a - 1$, for the mean model parameters, $\lambda_{\phi} = 1.026$ and 
$\lambda_{\chi} = 0.962$. 
We also show the evolution of 
$w_{\rm DE,CPL}$ for the Taylor-expanded 
CPL model.
In the case of the two-field quintessence model, $w_{\rm DE}$ is initially close to $-1$ and begins to deviate from $-1$ at low redshifts. This thawing behavior is analogous to that of a single-field exponential potential with slope $\lambda$ smaller than 1, but in the present case it is driven by the cooperative dynamics of two exponential potentials with $\lambda_{\phi} \sim \lambda_{\chi} \sim 1$. 
For the mean values of $\lambda_{\phi}$ and $\lambda_{\chi}$, the Universe enters the epoch of cosmic acceleration ($w_{\rm eff} < -1/3$) at $z < 0.65$. 
Unlike the evolution of $w_{\rm DE,CPL}$ in the CPL parametrization, the DE equation of state in the two-field quintessence model does not cross the cosmological-constant boundary ($w_{\rm DE} = -1$).

As shown in Table~\ref{tab:MCMC_bf}, the two-field quintessence model demonstrates a statistically notable improvement over the flat $\Lambda$CDM model, with a $\chi^2$ difference of $\Delta \chi^2 \sim -6.77$. Furthermore, the Bayes factor difference between the two models is $\Delta \ln B \sim 4$, indicating a moderate level of statistical preference for the two-field quintessence model. The CPL parametrization of $w_{\rm DE}$ yields a larger Bayes factor difference ($\Delta \ln B \sim 6.8$) relative to that of the $\Lambda$CDM model, which is attributed to its ability to realize a phantom-divide crossing. 
However, the preference for the two-field quintessence model over the $\Lambda$CDM model implies that current observational data support dynamical DE, even though the variation of $w_{\rm DE}$ is restricted to the range 
$w_{\rm DE} > -1$. 
Moreover, the central values of $\lambda_{\phi}$ and $\lambda_{\chi}$ inferred from the data are close to 1, a feature that is consistent with theoretical predictions from higher-dimensional theories.

\section{Conclusions}
\label{sec:conclusions}

In this paper, we performed a fully Bayesian analysis of the two-field quintessence model with exponential potentials motivated by higher-dimensional theories. We utilized the latest BAO data from the DESI collaboration, together with the Planck CMB and DES Y5 SnIa data. Using our analysis pipeline, we evaluated the goodness of fit of the two-field quintessence model and compared it with that of the $\Lambda$CDM model. We found that the two-field model provides a much better fit to the data, with a log Bayes factor difference of 
$\Delta \ln B \sim 4$, indicating moderate evidence against the $\Lambda$CDM scenario.

We find that the slopes of the exponential potentials are observationally constrained to be $\lambda_{\phi} \sim \lambda_{\chi} \sim 1 \pm 0.55$ 
at the 68.3\,\% credible level. 
Since values of $\lambda_{\phi}$ and $\lambda_{\chi}$ of order 
unity are consistent with the data, this result is theoretically appealing from the perspective of higher-dimensional theories, in particular the swampland conjecture (SC) \cite{Ooguri:2018wrx}. 
Although single-field quintessence is not ruled out by current observations, its slope is constrained to be 
$\lambda \sim 0.5\text{--}0.7$ \cite{Akrami:2025zlb, Alestas:2024gxe,Bhattacharya:2024hep}, 
which is smaller than the value suggested by the SC. 
In contrast, by considering two scalar fields, the SC can be simultaneously satisfied for each field, 
as the data favor $\lambda_\phi \sim \lambda_\chi \sim 1$. 
Moreover, owing to the assisted dynamics of two scalar fields with exponential potentials, the effective slope $\lambda_{\rm eff}$ can take values smaller than unity. 
In particular, we obtain the observational bound $\lambda_{\rm eff} \simeq 0.5 \pm 0.25$ at the 68.3\,\% credible level.

It should be noted that, while quintessence models with exponential potentials generally exhibit scaling/tracking behavior for $\lambda^2>4$ \cite{Copeland:1997et, Guo:2003eu}, we find that this theoretically attractive feature appears to be disfavored by the data, even with the additional freedom provided by two potentials.

Still, the two-field quintessence model offers a physically 
well-motivated and compelling mechanism that achieves a better fit to the data than $\Lambda$CDM.
We emphasize that this result is obtained without resorting to unphysical Taylor
expansions of $w_{\rm DE}(z)$ or ad~hoc parametrizations of 
$w_{\rm DE}(z)$.
Hopefully, forthcoming data from current surveys such as \textit{Euclid} will provide an independent test of the robustness of the DESI BAO measurements and shed further light on the nature of dynamical DE.

While the two-field quintessence model studied in this paper has a DE equation of state restricted 
to the range $w_{\rm DE} \geq -1$, it is possible to construct
scalar-tensor DE models that allow the crossing of the phantom 
divide from $w_{\rm DE} < -1$ to $w_{\rm DE} > -1$ \cite{Ye:2024ywg,Wolf:2025jed,Tsujikawa:2025wca,Wolf:2025acj,Tsujikawa:2026xqm}. The same can also happen with models of quintessence with an exponential potential, with a non-minimal coupling to gravity\cite{SanchezLopez:2025uzw,Bedroya:2025fwh}. 

Since these models possess explicit Lagrangians, as in the case of two-field quintessence, their cosmological dynamics are fully determined once the model parameters and initial conditions are specified. It would be of interest to investigate whether an MCMC analysis of such theoretical models, without resorting to a phenomenological parametrization of $w_{\rm DE}(z)$, can provide robust evidence for dynamical DE with future high-precision data, particularly concerning the phantom-divide crossing.

~\newline \paragraph{Code availability:}
The numerical codes used in our analysis, along with the MCMC chains, will be made publicly available after publication at \url{https://github.com/snesseris/two_field_quintessence}  \newline

\begin{figure*}[!t]
\includegraphics[width=\textwidth]{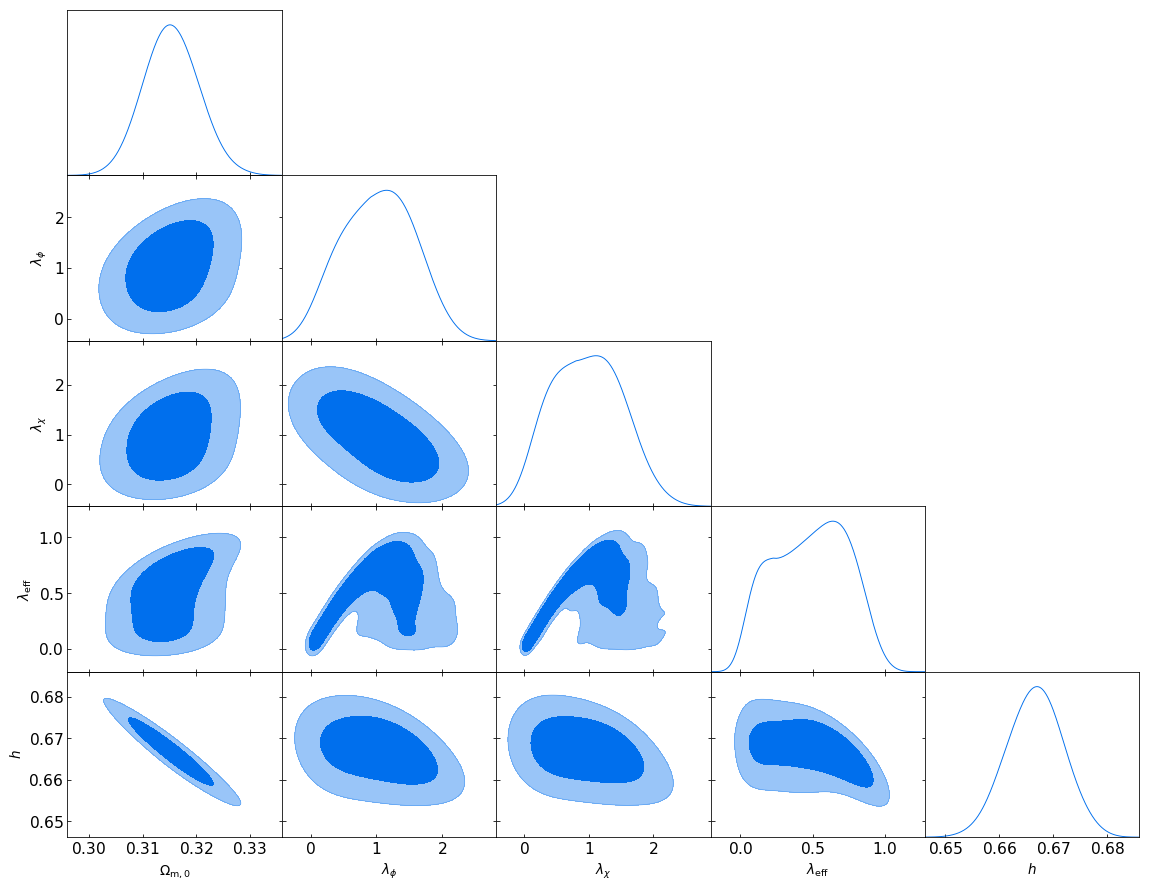}
\caption{
The one-dimensional marginalized posteriors and the two-dimensional contours are shown for the parameters of the two-field quintessence model, including the effective slope $\lambda_{\mathrm{eff}}$. 
The 68.3\,\% (inner) and 95.5\,\% (outer) credible intervals are obtained using CMB, DESI DR2, and DES Y5 data.
\label{fig:triangle}
}
\end{figure*}
%

\section*{Acknowledgements}

The authors thank S.~Kuroyanagi for useful discussions at an early stage of this work, and Y.~Akrami for interesting discussions on the analysis. 
They also acknowledge the use of the  \textit{San Calisto} supercomputer at the IFT. 
SN, GA, MC and IO thank Waseda University for warm hospitality at the beginning of the work and also acknowledge support from the research projects PID2021-123012NB-C43, PID2024-159420NB-C43, and the Spanish Research Agency (Agencia Estatal de Investigaci\'on) through the Grant IFT Centro de Excelencia Severo Ochoa No CEX2020-001007-S, funded by MCIN/AEI/10.13039/501100011033. 
GA's  research is supported by the Spanish Atracci\'on de Talento contract no.~2019-T1/TIC-13177 and 2023-5A/TIC-28945 granted by the Comunidad de Madrid. 
MC acknowledges support from the ``Ram\'on Areces'' Foundation through the ``Programa de Ayudas Fundaci\'on Ram\'on Areces para la realizaci\'on de Tesis Doctorales en Ciencias de la Vida y de la Materia 2023'' and the hospitality of Max Planck Institute for Gravitational Physics (Albert Einstein Institute) during the period in which part of this work was completed. 
IO is also supported by the fellowship LCF/BQ/DI22/11940033 from ``la Caixa" Foundation (ID 100010434).
This work is partially funded by the European Commission - NextGenerationEU, through Momentum CSIC Programme: Develop Your Digital Talent. Numerical calculations have been performed on the Hydra cluster at IFT. We acknowledge HPC support by Emilio Ambite, staff hired under the Generation D initiative, promoted by Red.es, an organisation attached to the Spanish Ministry for Digital Transformation and the Civil Service, for the attraction and retention of talent through grants and training contracts, financed by the Recovery, Transformation and Resilience Plan through the European Union's Next Generation funds. 
ST thanks the members of IFT for their warm hospitality during the visit.
ST is supported by JSPS KAKENHI Grant Number 22K03642 and Waseda University Special Research Projects (Nos.~2025C-488 and 2025R-028).

\appendix 
\section{
The triangle plots \label{sec:app1}}
In Fig.~\ref{fig:triangle}, we show the full triangle plot with 
the 1D marginalized posteriors and the 2D confidence regions 
for all the parameters of the double quintessence model.

\begin{figure}[!t]
\includegraphics[width=\columnwidth]{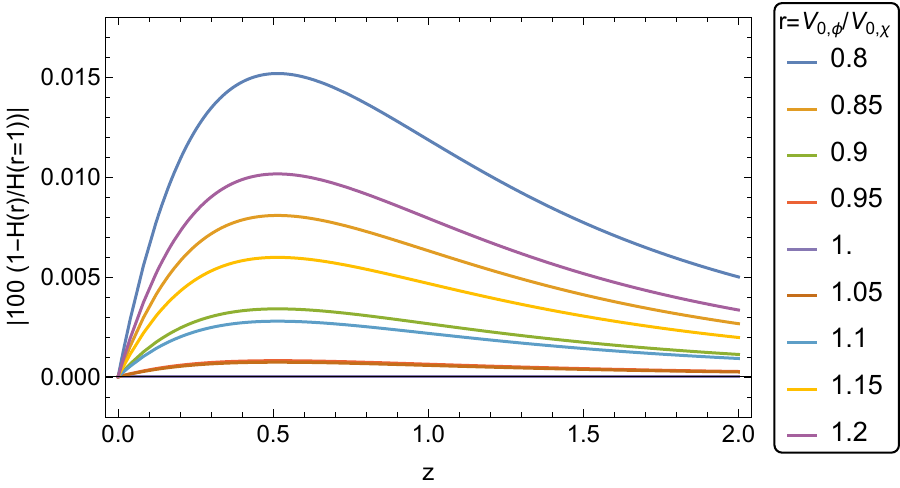}
\caption{The absolute value of the variation of the Hubble parameter as a function of redshift, for different values of the parameter 
$r\equiv V_{0\phi}/V_{0\chi}$. As can be seen, even large variations of $r$, of the order of $\sim 20\,\%$, only cause small changes in the Hubble parameter, less than $\sim 0.015\,\%$ with respect to 
the $r=1$ case.
\label{fig:ratio}}
\end{figure}
\begin{figure*}[!t]
\includegraphics[width=\textwidth]{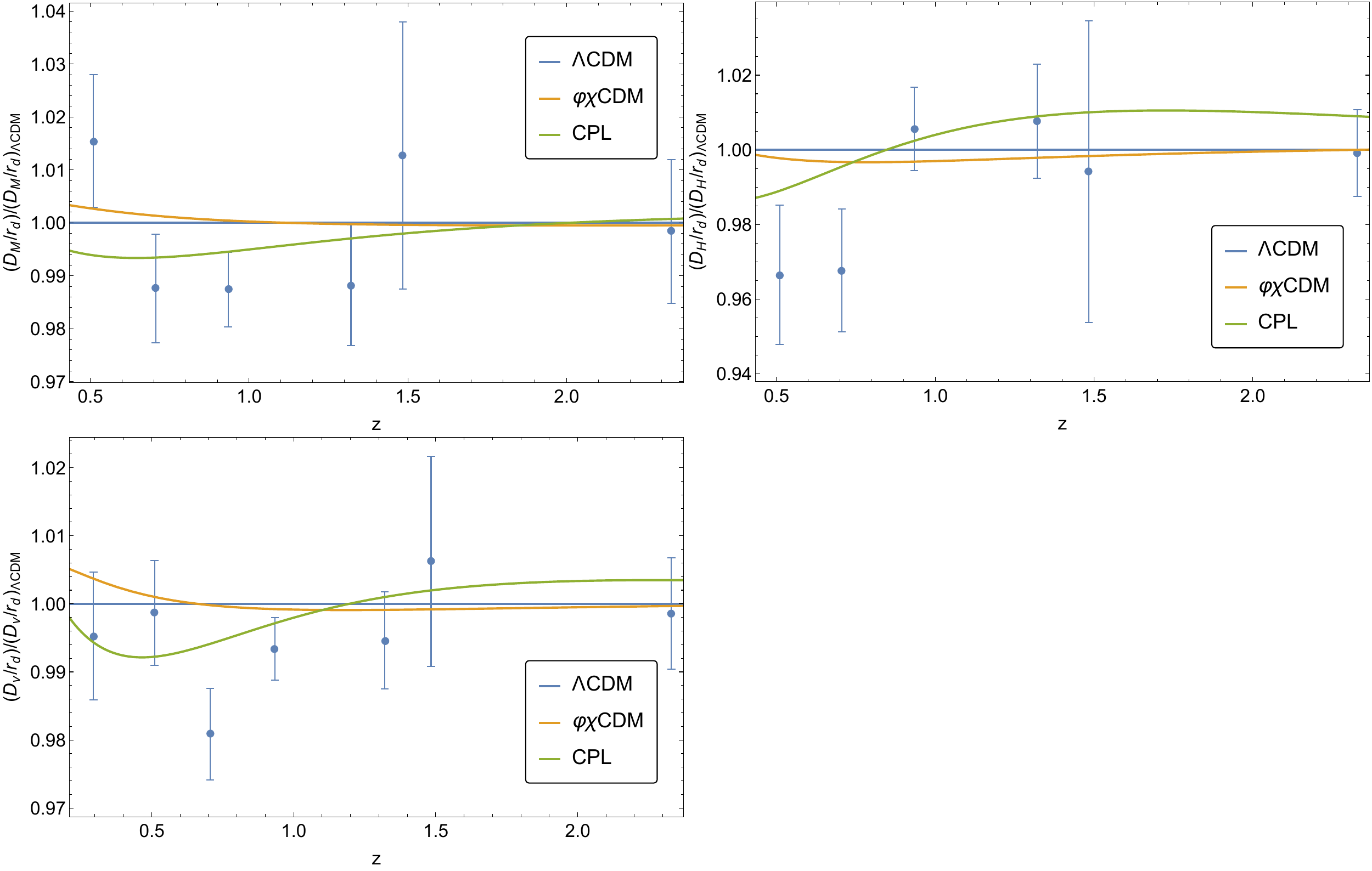}
\caption{We show the DESI DR2 measurements of the comoving transverse distance $D_{M}/r_d$ (left), the Hubble distance $D_{H}/r_d$ (right), 
and the volume-averaged distance $D_{V}/r_d$ (bottom), each normalized by the corresponding $\Lambda$CDM best-fit model. 
The error bars reflect the propagated DESI covariance, while the curves show the theoretical predictions for the $\phi\chi$CDM and CPL models, normalized in the same manner.
\label{fig:distances}
}
\end{figure*}

\section{
Different potential amplitudes \label{sec:app2}}

Here we explore the possibility of unequal amplitudes 
in the potentials of the two scalar fields. 
We study the dependence of the expansion history on
$r\equiv V_{0\phi}/V_{0\chi}$, by varying $r$ and applying 
a shooting method to $V_{0\phi}$, as the number of free 
parameters is increased.

As can be seen in Fig.~\ref{fig:ratio}, even relatively 
large variations in $r$ at the level of $\mathcal{O}(20\,\%)$ 
lead to at most a $\sim 0.015\,\%$ change in the Hubble parameter relative to the equal-amplitude case $(r=1)$. 
On average, the deviation is only at the level of 
$\sim 0.01\,\%$ when the full redshift range 
$z \in [0,2]$ is considered.

\section{Normalized DESI BAO distance functions \label{sec:app3}}

In this Appendix, as shown in Fig.~\ref{fig:distances}, we present the DESI BAO distance measurements expressed as ratios with respect to our best-fit $\Lambda$CDM model. 
The data points correspond to the DESI DR2 measurements, 
translated into $D_{M}(z)/r_d$, $D_{H}(z)/r_d$, and $D_{V}(z)/r_d$, with uncertainties propagated consistently.

As can be seen, the fact that the CPL parametrization provides a better fit to the data may indeed be attributed to its phantom crossing around $z \sim 0.5$. Since it is not statistically sound to assess the quality of the fit ``by eye'', we performed a Bayesian analysis of the 
two-field quintessence model in the main text (see Table~\ref{tab:MCMC_bf} and Sec.~\ref{sec:results}). 
We find that this model provides a better fit to the data than $\Lambda$CDM, but performs worse than the CPL parametrization.

\bibliographystyle{apsrev4-1}
\bibliography{Bibliography}

\end{document}